\documentclass[a4paper,11pt]{article}
\usepackage{jcappub} 
\makeatletter
\gdef\@fpheader{ }
\makeatother

\usepackage[export]{adjustbox}

\renewcommand{\b}[1]{{\bf #1}}
\newcommand{\bs}[1]{{\boldsymbol #1}} 

\newcommand{\nc}{\newcommand}
\nc{\ba}{\begin{eqnarray}}
\nc{\ea}{\end{eqnarray}}

\title{Boundary Crossing in Stochastic Inflation with Critical Number of Fields}

\author[a,b]{Mahdiyar~Noorbala}
\author[b,c]{Hassan~Firouzjahi,}

\affiliation[a]{Department of Physics, University of Tehran, P.O.~Box 14395-547, Tehran, Iran}
\affiliation[b]{School of Astronomy, Institute for Research in Fundamental Sciences (IPM), P.O.~Box 19395-5531, Tehran, Iran}
\affiliation[c]{Department of Physics, Faculty of Basic Sciences, University of Mazandaran, P.O.~Box 47416-95447, Babolsar, Iran}

\emailAdd{mnoorbala@ut.ac.ir}
\emailAdd{firouz@ipm.ir}

\begin{document}

\abstract{We study boundary crossing probability in the context of stochastic inflation.  We prove that for a generic multi-field inflationary potential, the probability that the inflaton reaches infinitely far regions in the field space is critically dependent on the number of fields, being nonzero for more than two fields, and zero otherwise.  We also provide several examples where the boundary crossing probability can be calculated exactly, most notably, for a particular landscape of a two-field model with a multi-well potential.}

\maketitle

\section{Introduction}\label{sec:introduction}

Inflation~\cite{Starobinsky:1980te, Sato:1980yn, Guth:1980zm, Linde:1981mu, Albrecht:1982wi, Linde:1983gd} is arguably the most prominent theory for describing the dynamics of the early universe, addressing the flatness and horizon problems of the standard cosmology, as well as explaining the origin of the primordial perturbations that are seen in the CMB and that later seed the large scale structures~\cite{Mukhanov:1981xt, Hawking:1982cz,  Starobinsky:1982ee, Guth:1982ec, Bardeen:1983qw, Starobinsky:1979ty}.

The simplest models of inflation are based on a scalar field dynamics which slowly rolls on top of a flat potential. The quantum fluctuations of the light inflaton field are continuously generated in each Hubble patch which are subsequently  stretched to superhorizon scales, to seed the CMB perturbations and the large scale structures.  While the simple single field slow-roll inflation scenarios are well consistent with cosmological observation, however one can also look for more complicated scenarios, such as multiple field models.  The basic predictions of models of inflation are that the primordial perturbations to be nearly scale invariant, nearly adiabatic and nearly Gaussian, which are in good agreement with observations \cite{Akrami:2018odb, Ade:2015lrj}.

The stochastic formalism~\cite{Vilenkin:1983xp, Starobinsky:1986fx, Rey:1986zk, Nakao:1988yi, Sasaki:1987gy, Nambu:1987ef, Nambu:1988je,  Kandrup:1988sc,  Nambu:1989uf, Mollerach:1990zf, Linde:1993xx, Starobinsky:1994bd, Kunze:2006tu, Prokopec:2007ak, Prokopec:2008gw, Tsamis:2005hd, Enqvist:2008kt, 
Finelli:2008zg, Finelli:2010sh, Garbrecht:2013coa, Garbrecht:2014dca, Burgess:2014eoa, 
Burgess:2015ajz, Boyanovsky:2015tba,  Boyanovsky:2015jen, Fujita:2017lfu} is based on coarse-graining the inflaton field's fluctuations  by discarding the short-wavelength modes and considering their effect as a classical noise on the long-wavelength modes.  The long-wavelength field is then shown to obey, under the slow roll condition, the Langevin equation
\begin{equation}
\frac{d\phi}{dN} + \frac{V_{, \phi}}{3H^2} = \frac{H}{2\pi} \xi(N),
\end{equation}
where $N$ is the number of $e$-folds and $\xi$ is the white Gaussian noise, satisfying
\ba
\label{xi-noise}
\big \langle \xi\left(N\right)\big \rangle = 0 \, , \quad \quad 
\big \langle \xi\left(N\right)\xi\left(N'\right)\big \rangle =\delta\left(N-N'\right) \, .
\ea 

This approach has been used to study eternal inflation~\cite{Vilenkin:1983xq, Linde:1986fd, Aryal:1987vn, Linde:1993xx, Linde:1996hg}, and more recently, in combination with the $\delta N$ formalism to study the correlation functions of the curvature perturbations~\cite{Fujita:2013cna, Fujita:2014tja, Vennin:2015hra, Vennin:2016wnk, Assadullahi:2016gkk, Grain:2017dqa, Noorbala:2018zlv, Firouzjahi:2018vet, Pattison:2019hef}. The $\delta N$ formalism \cite{Sasaki:1995aw, Sasaki:1998ug, Lyth:2004gb, Wands:2000dp, Lyth:2005fi} is based on the separate Universe approach in which the background expansion histories of the nearby Universes are modified in the presence of the superhorizon perturbations. With this picture in mind, one expects that the stochastic $\delta N$ formalism to be the right tool to study the dynamics of the superhorizon perturbations and calculate various correlation functions such as the curvature perturbation power spectrum and bispectrum \cite{ Vennin:2015hra, Vennin:2016wnk, Assadullahi:2016gkk}.

One of the quantities that the stochastic formalism makes easy to calculate is the boundary crossing probability, which is the probability that the inflaton crosses a particular surface (like the reheating surface, or a large-field cutoff surface) in the field space.  It was shown in Ref.~\cite{Assadullahi:2016gkk} that the large-field exploration probability is critically sensitive to the number of fields $D$, vanishing for $D\leq2$ and being nonzero for $D>2$.  The analysis of Ref.~\cite{Assadullahi:2016gkk} is based on a special class of potentials, called $v(r)$-potentials, which have a spherical symmetry in the field space.  One of the main results of this paper is to generalize that result to generic potentials, which we call $v(\phi)$-criticality, as opposed to the $v(r)$-criticality of Ref.~\cite{Assadullahi:2016gkk}.  This is done in section~\ref{sec:critical}.  In addition, we provide several examples in which the boundary crossing probability is explicitly calculated.

This paper is organized as follows: In section~\ref{sec:review} we review the boundary crossing probability and the basic equation governing it.  In section~\ref{sec:critical} we define what we mean by the critical behavior of boundary crossing probability, and then prove that it happens in $D=2$.  In section~\ref{sec:2d} we consider the two-field problem and offer two methods for computing the boundary crossing probability; we also provide several examples.  Sections~\ref{sec:critical} and \ref{sec:2d} are independent and can be read separately.  Finally, we summarize and conclude in section~\ref{sec:summary}.  Some detailed calculations are presented in appendices~\ref{app:harmonic} and \ref{app:image}.

\section{Review of Boundary Crossing Probability}\label{sec:review}

In the context of inflationary cosmology, one is often interested in finding out when the inflaton reaches a particular region of its field space.  This happens, for example, when one studies the probability of reheating.  Indeed, in most models, reheating takes place on a particular surface in the field space, namely, the reheating surface.  In a broader context, in a landscape of vacua, one often studies the probability of the inflaton reaching some particular vacuum.  Again, this can be regarded as the probability to cross a surface surrounding that vacuum.  In all of these examples, the central concept is ``boundary crossing probability'', so there is reasonable motivation for studying it and developing tools to calculate it.

\begin{figure}
\centering
\includegraphics[scale=.5]{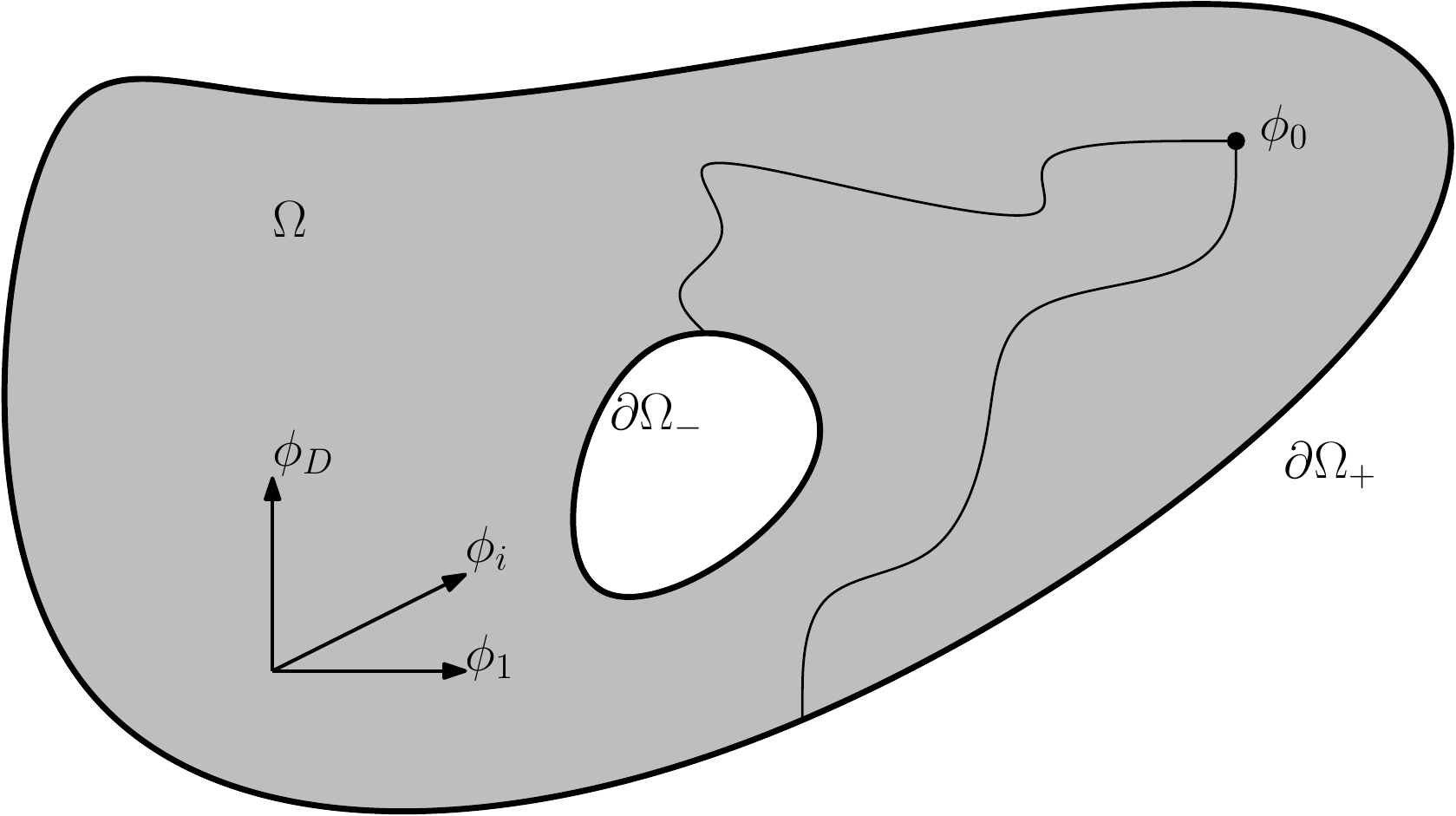}
\caption{The domain $\Omega$ and its boundaries $\partial\Omega_\pm$ in the field space.  Two trajectories, both starting from $\phi_0$, are drawn, one hitting $\partial\Omega_-$ and the other hitting $\partial\Omega_+$.}
\label{fig:paths}
\end{figure}

In this section we introduce the boundary crossing probability and the master equation it satisfies.  By boundary, we mean the boundary of a region $\Omega$ of the field space.  These are scalar fields $\phi_i$ (for $i=1,\ldots,D$) that can play the role of the inflaton under the potential 
\begin{equation}
V(\phi_1, \ldots, \phi_D) = 24\pi^2 M_\text{Pl}^4 v(\phi)  > 0,
\end{equation}
where $M_\text{Pl}$ is the reduced Planck mass, $v(\phi)$ is dimensionless potential 
and we have denoted by $D$ the dimension of the field space.  Throughout this paper and unless stated otherwise, we assume the slow roll conditions on the potential, and consider a connected region $\Omega$ that has two boundaries: an outer boundary $\partial\Omega_+$, and an inner one $\partial\Omega_-$.  Let $\phi_0\in\Omega$ be the starting point of inflaton in field space, as in Figure~\ref{fig:paths}.  Our goal is to study the probability $p_\pm$ that the inflaton hits $\partial\Omega_\pm$. The boundaries, for example, can be the reheating surface or the surface representing the UV cutoff of the large field models  etc. 

Defining  $p_\pm (\phi)$ as the probability that, starting from $\phi$, the inflaton crosses $\partial\Omega_\pm$,   it is shown in Ref.~\cite{Assadullahi:2016gkk} that
\begin{equation}\label{master}
\left( v \partial_i\partial_i - \frac{\partial_i v}{v} \partial_i \right) p_\pm(\phi) = 0,
\end{equation}
where $\partial_i = \partial/\partial\phi_i$.  The equation has to be solved with boundary conditions
\begin{equation}\label{boundary-conditions}
p_\pm(\partial\Omega_\pm)=1, \qquad p_\pm(\partial\Omega_\mp)=0.
\end{equation}
In the sequel we refer to Eq.~\eqref{master} as the \textit{master equation}.

There is a special class of potentials and boundary shapes for which the master equation can be solved analytically.  These so-called $v(r)$-potentials are functions of $r = \sqrt{\phi_1^2+\ldots+\phi_D^2}$.  In addition, the domain is annular, namely, $\Omega = \{ \phi | r_- \leq r \leq r_+ \}$.  Then it is easy to see that
\begin{equation}\label{vr-solution}
p_\pm (\phi) = \pm \displaystyle \dfrac{ \displaystyle \int_{r_\mp}^{r(\phi)} r'^{1-D} e^{-\frac{1}{v(r')}} d r'}{\displaystyle\int_{r_-}^{r_+} r'^{1-D} e^{-\frac{1}{v(r')}} dr'}.
\end{equation}
It was noted in Ref.~\cite{Assadullahi:2016gkk} that for $D\leq2$ and $\lim_{r\to\infty}v\neq0$, the solutions have an $r_+\to\infty$ limit
\begin{equation}\label{vr-fake-sol}
p_+ (\phi) = 1 - p_- (\phi) =
\begin{cases}
1 & \phi\in\partial\Omega_+, \\
0 & \phi\in\operatorname{int}(\Omega), \\
0  & \phi\in\partial\Omega_-,
\end{cases}
\end{equation}
where $\operatorname{int}(\Omega) = \Omega - \partial\Omega$ is the interior of $\Omega$.  This is the basis of the observation that $D=2$ is the critical dimension for the boundary crossing probability.  In the next section we generalize this result (which was proven for $v(r)$-potentials only in Ref.~\cite{Assadullahi:2016gkk}) to generic potentials.

\section{Critical Behavior of the Boundary Crossing Probability}\label{sec:critical}

In this section, we study the behavior of the boundary crossing probability $p_\pm$ as we vary the dimension $D$.  The potential $v(\phi)$ itself can have explicit dependence on $D$ (as in, e.g.,  $v = \phi_1+D\phi_2$).  There is also an implicit $D$-dependence of $v$ due to the fact that it is a $D$-variable function (as in the $v(r)$-potentials).  So let us assume a given family of potentials indexed by $D$ (which for the most part remains arbitrary), and study the solutions of Eq.~\eqref{master} as $D$ is varied.

Our motivation is to generalize
\begin{quote}
\textbf{$v(r)$-criticality:} Let $v(\phi)>0$ be a function of $r = \sqrt{\phi_1^2+\ldots+\phi_D^2}$ such that $\lim_{r\to\infty}v\neq0$, and let $\Omega = \{ \phi | r_- \leq r \leq r_+ \}$.  Then in the limit $r_+\to\infty$, for every $\phi$ in the interior of $\Omega$ we have $p_+(\phi)=0$ if $D\leq2$, and $p_+(\phi)\neq0$ if $D>2$.
\end{quote}
to the stronger statement
\begin{quote}
\textbf{$v(\phi)$-criticality:} Let $v(\phi)>0$ be such that $\lim_{|\phi|\to\infty}v\neq0$, and let $\Omega$ be an unbounded domain topologically equivalent to the exterior of the $(D-1)$-sphere $S^{D-1}$ in $\mathbb{R}^D$.  Then for every $\phi$ in the interior of $\Omega$ we have $p_+(\phi)=0$ if $D\leq2$, and $p_+(\phi)\neq0$ if $D>2$.
\end{quote}
The aforementioned results state that there is a critical behavior in $p_+(\phi)$ at $D=2$.  If $D>2$ there is always a non-zero probability for the inflaton to reach infinity ($r=\infty$).  But if $D\leq2$, there is absolutely zero probability for the inflaton to reach infinity no matter where it starts from.  The first statement ($v(r)$-criticality) is a direct consequence of Eq.~\eqref{vr-fake-sol} and is stated and proven in Ref.~\cite{Assadullahi:2016gkk}.  The second statement relaxes the restrictive assumption of radial symmetry on the potential and the domain, and is one of the results we intend to prove in this paper.

To make our statement concrete, let us focus first on the domain $\Omega$ in the field space.  $v(r)$-criticality assumes an unbounded $\Omega$.  This turns out to be essential in our generalized version, namely $v(\phi)$-criticality, too.  Let us see how this works out.  To begin, we note that Eq.~\eqref{master} is an elliptic equation to which the maximum principle applies, i.e., the functions $p_\pm$ take on their maximum and minimum values on the boundary of $\Omega$.\footnote{Here's a partial proof: Suppose that $\phi_0$ is a local maximum.  Then $\partial_i p=0$ at $\phi_0$ and hence Eq.~\eqref{master} implies that the trace of Hessian vanishes, namely $\partial_i \partial_i p=0$ at $\phi_0$.  This is a contradiction with the fact that at a local maximum the Hessian is positive definite.  Actually, the positivity of the second derivative is only a sufficient condition (consider $y=x^4$); for a complete proof on bounded and unbounded domains see~\cite{Gilbarg}.}

Given the boundary conditions $p_\pm(\partial\Omega_\pm)=1$ and $p_\pm(\partial\Omega_\mp)=0$, this implies the strict inequality $0<p_\pm(\phi)<1$ for any $\phi$ in the interior of $\Omega$.  At first glance, this looks contradictory even to $v(r)$-criticality, as the latter asserts that $p_+=0$ for $D\leq2$.  The point is that, strictly speaking, \eqref{vr-fake-sol} is not a solution of the master equation at all, as it is discontinuous at $\partial\Omega_+$.  It is only the $r_+\to\infty$ limit of the solutions of Eq.~\eqref{master}.  In fact the master equation has no solution in $D\leq2$ on the unbounded domain $\Omega = \{ \phi | r_- \leq r \}$, which is an indication of the critical behavior.

The discussion of the preceding paragraph guides us how to proceed.  Whenever a solution to the master equation exists, the maximum principle guarantees that $p_\pm$ must smoothly interpolate between the values $0$ and $1$ on the boundaries without any interior point having $p_\pm=0,1$.  On the other hand, the existence of solutions to the Dirichlet problem of elliptic equations (like our master equation~\eqref{master}) on (regular) \textit{bounded domains} is well established~\cite{Gilbarg}.  We conclude that in order to have a critical behavior, we need to find situations where no solution exists, which can only happen on unbounded domains.  As we saw above, the limiting solution \eqref{vr-fake-sol} is not continuous and hence cannot be regarded as a proper solution.  We call such fake solutions ``discontinuous solutions'' and we observe that they are the symptoms of critical behavior.  Note that the boundary crossing probabilities $p_\pm$ on the unbounded domain are given by the discontinuous solutions, although they are not proper solutions to the master equation.  So they are fake when considered as solutions to the differential equation~\eqref{master}, but quite genuine when regarded as boundary crossing probabilities. 

In the sequel, our strategy will be to show that for generic potential $v$ and on unbounded domain $\Omega$, (i) the proper solution to the master equation exists for $D>2$ (Subsection~\eqref{ssec:d>2}), and (ii) $D\leq2$ can only admit discontinuous solutions (Subsection~\eqref{ssec:d<2}).  This will establish that the critical behavior occurs at $D=2$.

\subsection{$D>2$}\label{ssec:d>2}

In this subsection we construct a proper solution of the master equation~\eqref{master} for $D>2$.  Although we will not uphold to the golden standards of mathematical rigor, we still need some degree of rigor.  Otherwise, we could simply assume that a solution exists and presume that it is continuous; as we saw, however, the nature of the question under study requires to be a bit more careful.  We therefore begin by assuming a bounded domain $\Omega$ for which rigorous existence results exist in the standard mathematical textbooks; see for example~\cite{Taylor}.

The trick to proceed is to write the master equation \eqref{master} as 
\begin{equation}
v e^{-1/v} \partial_i \left( e^{1/v} \partial_i p \right) = 0,
\end{equation}
with $p$ being either of $p_\pm$.  This can be cast in the form of the Laplace equation.  To see this, define the following metric on the field space
\begin{equation}\label{metric}
ds^2 = e^{\omega(\phi)} \left( d\phi_1^2 + \ldots + d\phi_D^2 \right).
\end{equation}
Then the Laplacian reads
\begin{equation}
\nabla^2 p = \frac{1}{\sqrt{g}} \partial_i \left( \sqrt g g^{ij} \partial_j p\right) = e^{-D\omega/2} \partial_i \left( e^{(D/2-1)\omega} \partial_i p \right).
\end{equation}
So with the choice, 
\begin{equation}\label{omega}
\omega(\phi) = \frac{2}{(D-2)v(\phi)},
\end{equation}
valid for $D>2$, the master equation becomes $\nabla^2p=0$ in the curved space described by the conformally flat metric \eqref{metric}.\footnote{We note that the requirement $\lim_{|\phi|\to\infty}v\neq0$ implies that $\omega$ is finite.  Then the conformally flat metric~\eqref{metric} preserves boundedness of regions of space.  Thus it is unambiguous when we talk about the boundedness of $\Omega$.}  This means that $p$ is a harmonic function on the curved space.  Of course the Laplace equation is elliptic and harmonic functions satisfy the maximum principle too, even on unbounded curved spaces, so they take their extrema on the boundaries.  But we have to show that a proper continuous solution does exist to begin with.

We can explicitly construct the solution by invoking the Dirichlet Green's function $G$ of Laplacian for the region $\Omega$ on the curved manifold
\begin{equation}
\begin{aligned}
\forall\phi,\phi'\in\Omega &:& \nabla'^2 G(\phi,\phi') &= \delta^{(D)}(\phi'-\phi), \\
\forall\phi\in\Omega,\phi'\in\partial\Omega &:& G(\phi,\phi') &= 0.
\end{aligned}
\end{equation}
Green's theorem on a curved manifold is a straightforward generalization of the more familiar version on flat space
\begin{equation}
\int_\Omega \left( f \nabla^2 g - g \nabla^2 f \right) \sqrt{g} d^D\phi = \int_{\partial\Omega} \left( f \frac{\partial g}{\partial n} - g \frac{\partial f}{\partial n} \right) \sqrt{h} d^{D-1}\phi,
\end{equation}
where $\partial/\partial n$ is the normal derivative on the boundary pointing away from $\Omega$, and $h$ is the induced metric on $\partial\Omega$.  It follows that $G(\phi,\phi')=G(\phi',\phi)$.  Now choose $f(\phi')=p_-(\phi')$ and $g(\phi')=G(\phi,\phi')$, to get
\begin{equation}\label{p=int-d2-G}
p_-(\phi) = -\int_{\partial\Omega_-} \frac{\partial G(\phi,\phi')}{\partial n'} \sqrt{h(\phi')} d^{D-1}\phi' = -\int_{\operatorname{int}(\partial\Omega_-)} \nabla'^2 G(\phi,\phi') \sqrt{g(\phi')} d^D\phi', 
\end{equation}
where again $\partial/\partial n$ points away from the interior of $\partial\Omega_-$, and we have used divergence theorem in the last expression.

We note that the last integral in Eq.~\eqref{p=int-d2-G} has a simple interpretation in terms of an electrostatic analogy.  $G(\phi,\phi')$ plays the role of the electrostatic potential at the position $\phi'$ in space due to a negative point charge at $\phi$, in the presence of grounded conductor surfaces located at $\partial\Omega$.  The image charges should be placed outside $\Omega$, so in those regions $-\nabla'^2 G(\phi,\phi')$ is the density of image charges.  Therefore, $p_-(\phi)$ is equal to the total interior image charge located inside $\partial\Omega_-$ when a negative unit point charge is placed at $\phi$.\footnote{As a simple example, consider the point charge $q$ at a distance $r$ away from the center of a grounded sphere of radius $r_-$.  The image charge inside the sphere is $q'=-qr_-/r$~\cite{Jackson}, in complete agreement with $p_-(r)$ of Eq.~\eqref{vr-solution} when $D=3$, $r_+=\infty$, and $v=\rm constant$.}  Since the image charges are positive and split to be either inside $\partial\Omega_-$ or outside $\partial\Omega_+$, it is clear that as $\phi$ moves from $\partial\Omega_-$ to $\partial\Omega_+$, $p_-(\phi)$ continuously interpolates between 1 and 0.

\begin{figure}
\centering
\includegraphics[scale=.7]{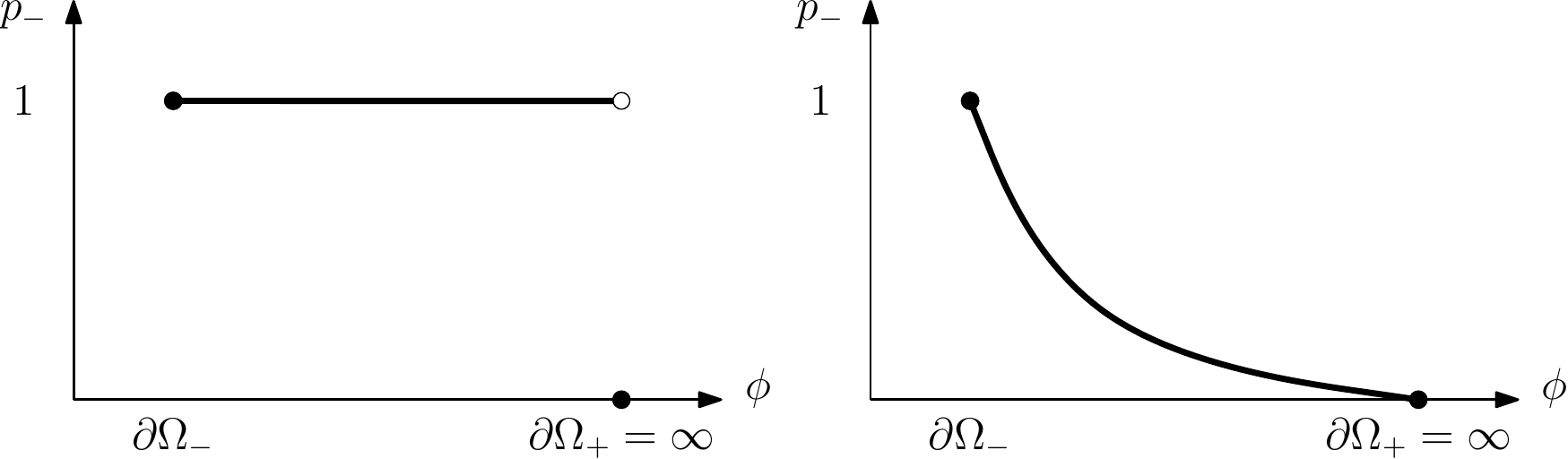}
\caption{Schematic plot of $p_-(\phi)$ in the limit of unbounded domain, for a discontinuous solution (left) and a continuous solution (right).  In both cases, the maximum value on the $\phi$ axis (i.e., the infinite boundary $\partial\Omega_+$) is placed at a finite distance from the origin.}
\label{fig:cont-vs-discont}
\end{figure}

It is now time to consider the case of unbounded $\Omega$ by taking the limit $\partial\Omega_+$ to infinity.  In order to prove that \eqref{p=int-d2-G} is a proper solution, we need to show that $p_-(\phi)$ remains continuous in this limit.  Recall that in a discontinuous solution, like Eq.~\eqref{vr-fake-sol}, $p_-(\phi)$ approaches 1 in this limit (except for $\phi\in\partial\Omega_+$, so that $p_-(\phi)$ is discontinuous at $\phi=\infty$; see Fig.~\ref{fig:cont-vs-discont}).  We show that, in contrast to Eq.~\eqref{vr-fake-sol}, the $p_-(\phi)$ in Eq.~\eqref{p=int-d2-G} approaches zero as $\phi$ goes to $\infty$, i.e.,
\begin{equation}\label{limlim}
\lim_{\phi\to\infty} \left[ \lim_{\partial\Omega_+\to\infty} p_-(\phi) \right] = 0.
\end{equation}

The key point to prove the above assertion is that the potential of point charge in $D$ dimensions has the asymptotic behavior $\propto 1/R^{D-2}$.  For a flat manifold $R$ is the Euclidean distance $|\phi-\phi'|$ between the source $\phi'$ and point of observation $\phi$.  For our conformally flat manifold, for which $\omega(\phi)$ approaches a constant 
\begin{equation}
\omega_\infty =  \lim_{|\phi|\to\infty} \omega(\phi) = \frac{2}{D-2} \lim_{|\phi|\to\infty} \frac{1}{v(\phi)},
\end{equation}
we have $R = e^{\omega_\infty/2} |\phi-\phi'|$.  Now let's push $\partial\Omega_+$ and then $\phi$ to infinity.  If $p_-(\phi)\neq0$ (contrary to our claim), then we have nonzero interior image charges and hence they contribute a nonzero potential at $\partial\Omega_-$.  This nonzero potential must be canceled by the other two sources of charge (the original charge at $\phi$, and the exterior charges beyond $\partial\Omega_+$ at infinity) to yield a grounded surface at $\partial\Omega_-$.  However, the latter are too far away to cancel the aforementioned nonzero potential (in other words, because of the $1/R^{D-2}$ decay, they can be made arbitrary small in the limit $\partial\Omega_+,\phi\to\infty$).  This shows that as $\partial\Omega_+$ is pushed to infinity, there are points $\phi$ at which $p_-(\phi)$ can be made arbitrarily small.  This is the promised result~\eqref{limlim}; thus $p_-(\phi)$ continuously interpolates from 1 to 0 over the unbounded domain $\Omega$.  Therefore, we have a proper continuous solution.

If all this looks trivial, then consider how the argument fails for $D=1,2$, where the potential of a point charge goes like $R$ and $\log R$, respectively.  It is precisely the decaying behavior $\propto 1/R^{D-2}$ of the potential at long distances that makes $p_-$ (as well as $p_+=1-p_-$) a proper solution to the master equation~\eqref{master} for $D>2$.  In fact, the same equation $\nabla^2 p=0$ does not have a proper solution on an unbounded 1- or 2-dimensional domain.  Of course, the results of this section leading to the metric~\eqref{metric} are not applicable in $D=2$, as the conformal factor~\eqref{omega} is undefined.  Our point is to emphasize the role of the decaying potential in $D>2$.  A more concrete illustration of this point in the special case of an annular region in flat space is presented in appendices~\ref{app:harmonic} and \ref{app:image}.

\subsection{$D\leq2$}\label{ssec:d<2}

We now turn attention to $D\leq2$, treating the $D=1$ and $D=2$ cases separately.  Our goal is to prove that a proper solution does not exist on an unbounded domain and that only discontinuous solutions exist.

When $D=1$, we can define a new variable $\varphi = \int e^{-1/v}d\phi$, which measures the distance in the metric \eqref{metric}, i.e., $ds=d\varphi$.  In terms of $\varphi$ the master equation~\eqref{master} reads $d^2p/d\varphi^2=0$, whose solution is $p=a\varphi+b$.  The constants $a$ and $b$ are to be determined by the boundary conditions~\eqref{boundary-conditions}.  Suppose $\partial\Omega_\pm = \{ \phi=\phi_\pm \} = \{ \varphi=\varphi_\pm \}$.  Then
\begin{equation}
p_\pm(\varphi) = \pm \frac{\varphi - \varphi_\mp}{\varphi_+ - \varphi_-}.
\end{equation}
For finite $\varphi_\pm$, this is a continuous solution.  But for $\varphi_+ = \infty$, which is the case for unbounded $\Omega$, we have $p_+=0$ and $p_-=1$ which are discontinuous at the boundaries.  We can express this in terms of the original variable $\phi$ as follows: If the boundaries are separated by infinite distance under the metric~\eqref{metric} (that is, $\int_{\phi_-}^{\phi_+} e^{-1/v}d\phi = \infty$), then the inflaton almost always ends up on the boundary which is a finite distance away from its starting point.  This establishes the assertion of $v(\phi)$-criticality for $D=1$.\footnote{It is important to note that even in $D\neq1$, the boundary conditions may render the $D$-dimensional problem to an effectively 1-dimensional one (for example, when $\partial\Omega_\pm$ and $v(\phi)$ depend only on one of the fields).  In those cases too (which we don't regard as genuine $D$-dimensional problems), we have $p_+=0$ when the boundaries are infinitely apart.}

When $D=2$, the assignment $\omega=2/(D-2)v$ breaks down and we no longer have harmonic functions on a curved manifold.  Nevertheless, we can proceed by proving that the space of solutions to the master equation~\eqref{master} is invariant under conformal transformations on the complex plane.  Let $z=\phi_1+i\phi_2$ and $z'=\phi'_1+i\phi'_2$ such that $z'(z)$ is a holomorphic function of $z$.  Define the metric
\begin{equation}
ds^2 = d\phi_1^2 + d\phi_2^2 = \left| \frac{\partial z}{\partial z'} \right|^2 \left( d\phi'^2_1 + d\phi'^2_2 \right).
\end{equation}
The quantity 
\begin{equation}
\nabla^2 p + g^{ij} \partial_i \frac{1}{v} \partial_j p
\end{equation}
is an invariant, whose value in the primed and unprimed coordinates is
\begin{equation}
\left| \frac{\partial z'}{\partial z} \right|^2 \left( \partial'_i \partial'_i p + \partial'_i \frac{1}{v} \partial'_i p \right) = \partial_i \partial_i p + \partial_i \frac{1}{v} \partial_i p.
\end{equation}
(The assumption of $z'(z)$ being a conformal transformation assures that the derivative $\partial z'/\partial z$ is nonzero.)  So if $p(z)$ is a solution of Eq.~\eqref{master}, then so is $p'(z') = p(z(z'))$.  Note that the region $\Omega$ as well as its boundary $\partial\Omega$ in the $z$-plane get mapped to $\Omega'$ and $\partial\Omega'$ in the $z'$-plane.  So we have two solutions to the master equation with two different boundary conditions.  We will later use this fact in subsection~\ref{ssec:conformal} to construct new solutions starting from a known one.  But for now let us use it to pursue our proof of nonexistence of proper solutions for $D=2$.

\begin{figure}
\centering
\includegraphics[scale=.9]{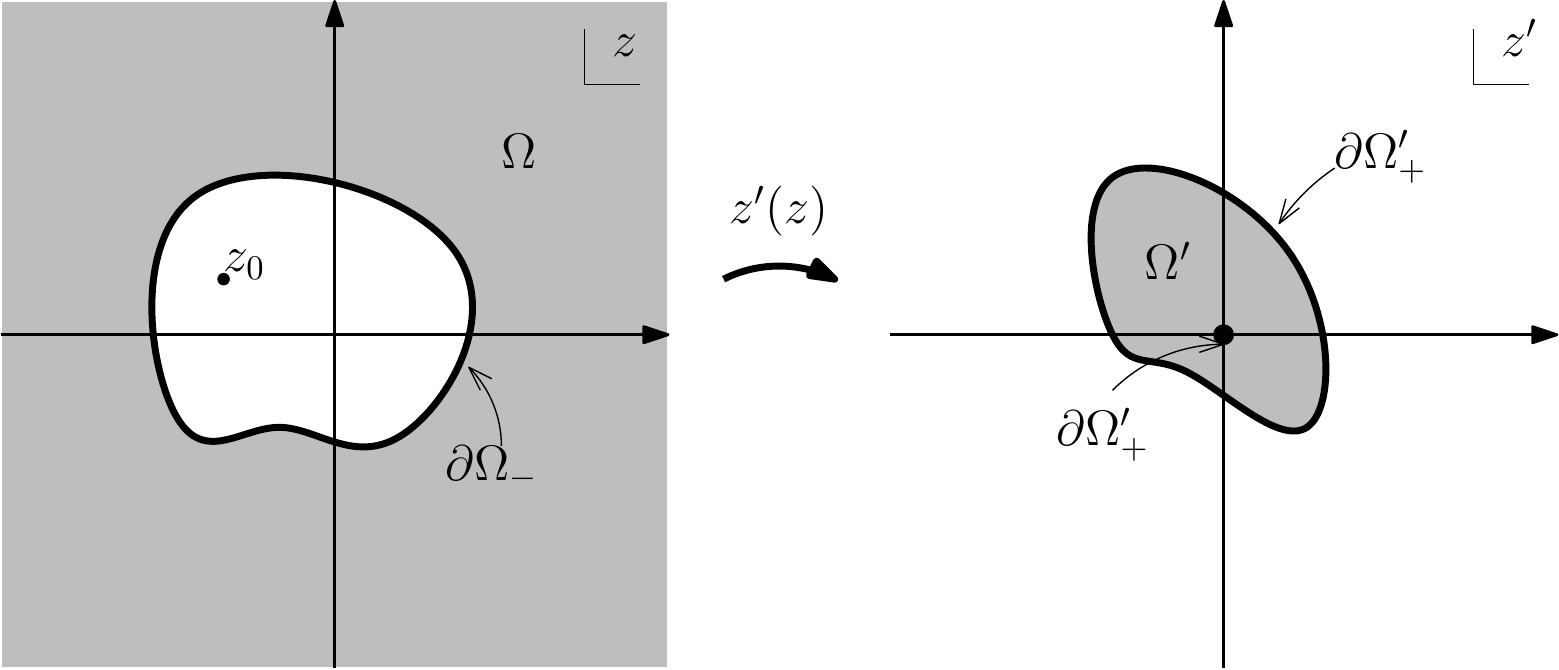}
\caption{The holomorphic map $z'(z)=1/(z-z_0)$ mapping the region $\Omega$ in the $z$-plane (left) to the region $\Omega'$ in the $z'$-plane (right).}
\label{fig:conformal-map}
\end{figure}

Take a point $z_0\in\operatorname{int}(\partial\Omega_-)$ outside the unbounded domain $\Omega$ (see figure~\ref{fig:conformal-map}).  The map $z'=1/(z-z_0)$ is holomorphic over $\Omega$ and maps it to the domain $\Omega'$ bounded by the outer boundary $\partial\Omega'_+$ (which is the image of $\partial\Omega_-$) with effectively no inner boundary ($\partial\Omega'_-=\{0\}$).  If a proper solution for $p_-$ exists on the original unbounded domain $\Omega$, then the map $z'(z)$ yields a proper solution $p'_+$ on the new bounded domain $\Omega'$.  Note that since we have $p_-(\infty)=0$ in the original problem, the new solution must have $p'_+(0)=0$.  But this is an overspecification, because the elliptic differential equation~\eqref{master} with Dirichlet boundary condition $p'_+=1$ on its sole boundary $\partial\Omega'_+$, already uniquely fixes the function $p'_+$ at all $z'\in\Omega'$ including $z'=0$.  In fact, the latter is a trivial problem whose solution is the constant function $p'(z')=1$.  This is in contradiction with $p'(0)=0$, so we conclude that a proper solution does not exist for $D=2$.

Put together, the results of this subsection show that, in contrast to $D>2$, there is no proper continuous solution of the master equation~\eqref{master} on an unbounded 1- or 2-dimensional domain.  Thus the only possible solution for $D\leq2$ is the discontinuous solution~\eqref{vr-fake-sol}.  This completes our proof of the $v(\phi)$-criticality and establishes $D=2$ as the critical dimension for general potentials.

\section{Boundary Crossing Probability in $D=2$}\label{sec:2d}

We have already seen in subsection~\ref{ssec:d<2} that the 2-dimensional problem can be dealt with by methods of complex analysis.  In this section we offer two methods for computing the boundary crossing probability of a two-field inflationary model (of course, on bounded domains, for which the solution is not trivially given by the discontinuous solution~\eqref{vr-fake-sol}).  We also provide some examples.

\subsection{Conformal Transformation}\label{ssec:conformal}

Our first method is based on the fact proved in subsection~\ref{ssec:d<2}: The space of solutions to the master equation is invariant under conformal transformations.  It enables us to take a $v(r)$-potential on the annulus $\Omega = \{ \phi | r_- \leq r \leq r_+ \}$, for which we know the solution $p$ from Eq.~\eqref{vr-solution}, pick an arbitrary conformal transformation $z'=f(z)$, and construct the new potential $v'$, domain $\Omega'$, and solution $p'$ as
\begin{equation}
v'(z') = v \left( f^{-1}(z') \right), \qquad \Omega' = f(\Omega), \qquad p'(z') = p \left( f^{-1}(z') \right),
\end{equation}
where
\begin{equation}\label{vr-solution-2d}
p_\pm (z) = \pm \displaystyle \dfrac{ \displaystyle \int_{r_\mp}^{|z|} e^{-\frac{1}{v(r)}} \frac{dr}{r}}{\displaystyle\int_{r_-}^{r_+} e^{-\frac{1}{v(r)}} \frac{dr}{r}}.
\end{equation}


We also need to show that slow rolling on the potential $v$ implies slow rolling on $v'$, since our starting point in deriving the master equation~\eqref{master} was the slow roll approximation.  It is sufficient to show that the derivative of $v'$ in an arbitrary direction is proportional to the derivative of $v$, since the latter is proportional to $\sqrt\epsilon v$.  Noting that $v'(z')=v(z)$, and that $z'$ is only a function of $z$ and not of $z^*$, we can write the chain rule
\begin{equation}
\frac{dv'(z')}{dz'} = \frac{\partial z}{\partial z'} \frac{dv(z)}{dz},
\end{equation}
which immediately implies the desired result.

The celebrated Riemann mapping theorem of complex analysis assures that any pair of simply connected domains can be conformally mapped to each other~\cite{Ahlfors}.  This is not true for the doubly connected domains we study.  So in general this method is not guaranteed to work, and when it does it may not be easy to find the conformal transformation $z'(z)$.  In fact, since contours of constant $v$ and constant $p$ coincide, such a conformal transformation can only generate harmonic potentials, which by definition have this property~\cite{Assadullahi:2016gkk}.  Below we demonstrate the method by two examples:

\subsubsection{Example: M\"obius transformation}

The M\"obius transformation
\begin{equation}
f(z) = \frac{az+b}{cz+d}, \qquad ad-bc\neq0
\end{equation}
is a combination of translation, rotation, scaling, and inversion off the unit circle.  It maps circles to circles (straight lines are regarded circles of infinite radius).

\begin{figure}
\centering
\includegraphics[width=.4\textwidth, trim=4.5cm 0 4.5cm 0, clip, valign=c]{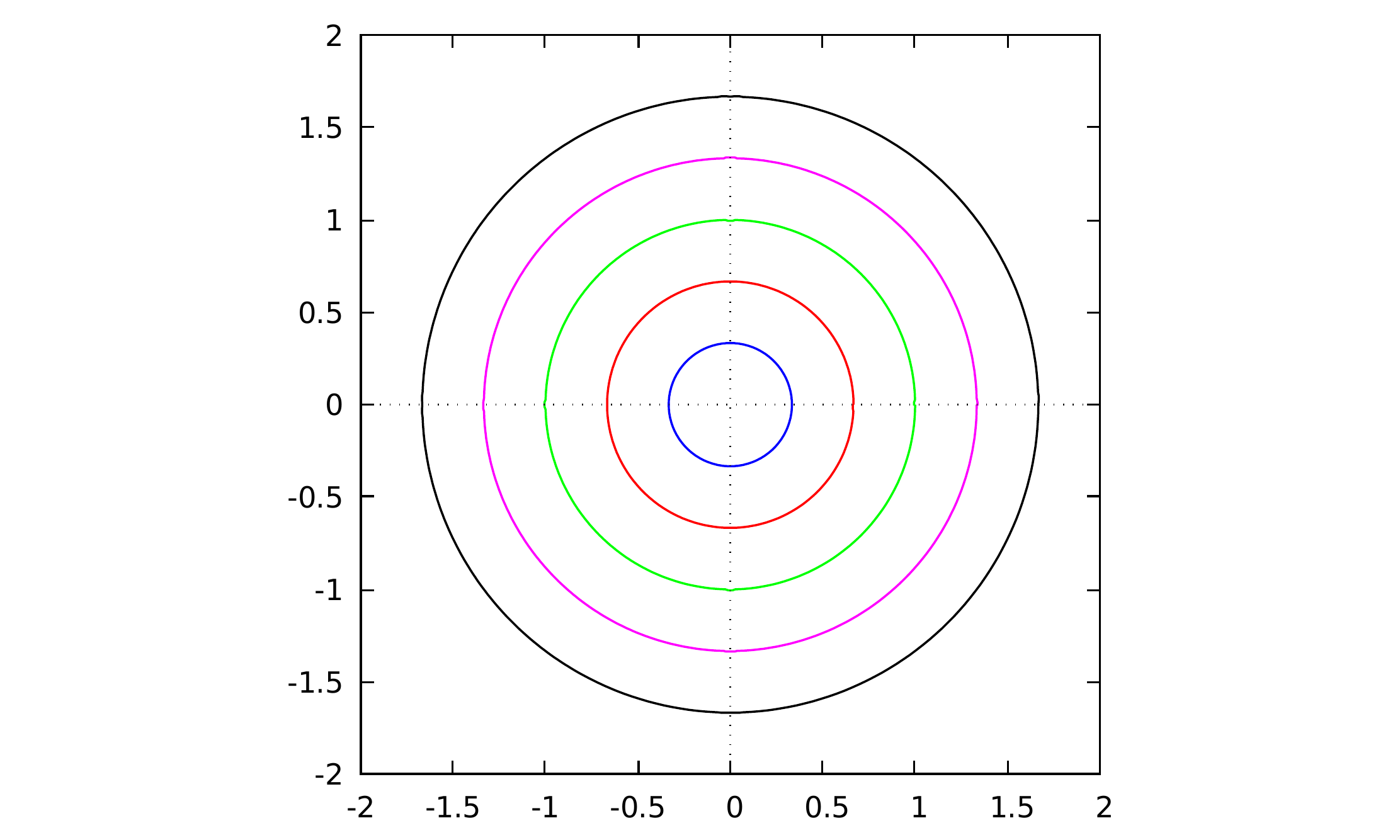}
\includegraphics[width=.55\textwidth, valign=c]{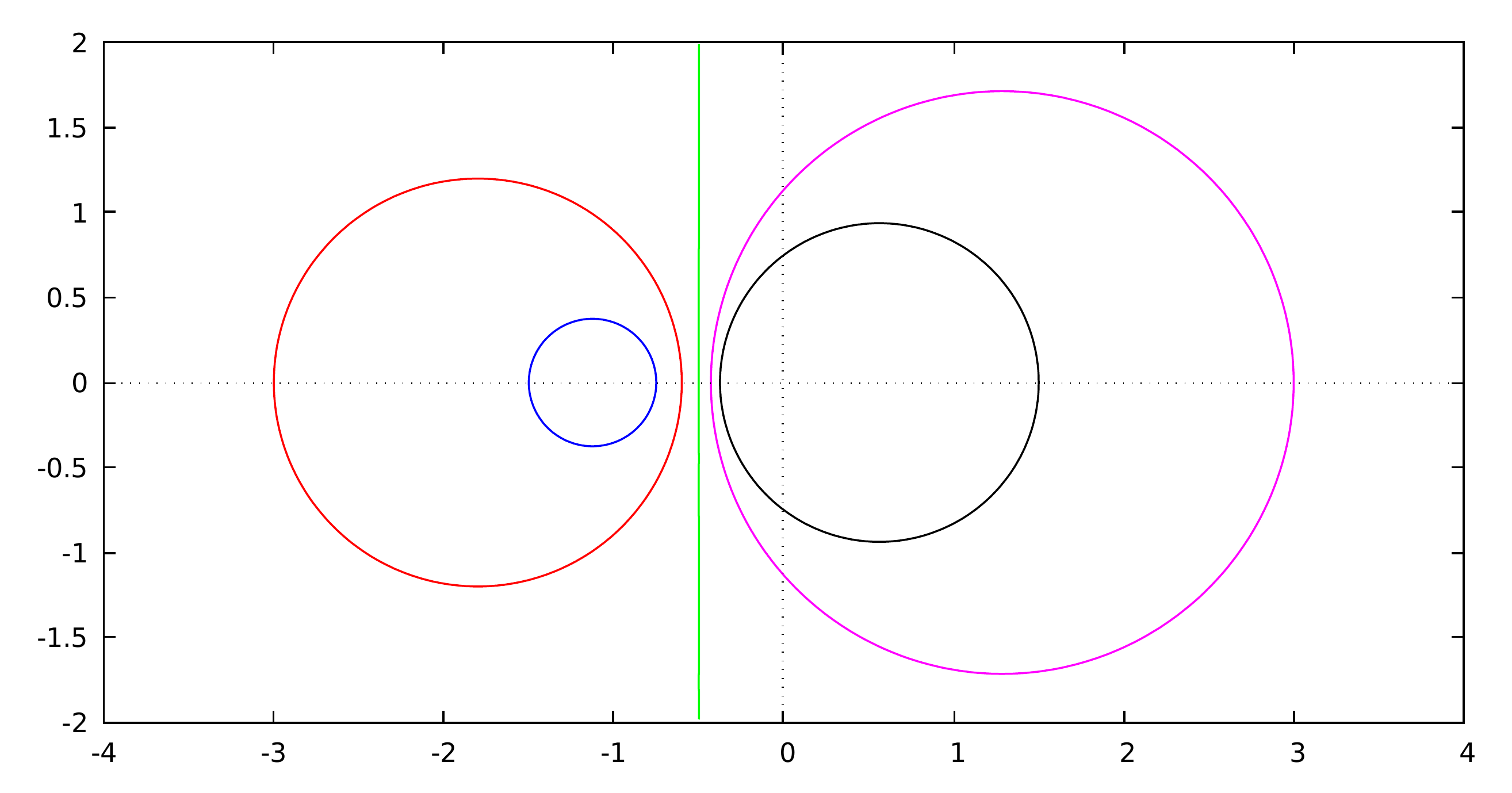}
\caption{The M\"obius transformation $z\mapsto z'=1/(z-1)$ mapping concentric circles with radii $n/3$ ($n=1,\ldots,5$) in the $z$-plane (left) to their images in the $z'$-plane (right).}
\label{fig:mobius}
\end{figure}

Let us apply $f(z)=1/(z-a)$ (with real $a$) to the $v(r)$-potential $v(|z|)$.  The contours of constant $v$ (circles centered at the origin in the $z$-plane) get mapped to contours of constant $v'$ in the $z'$-plane.  These are depicted in figure~\ref{fig:mobius}.  We can write
\begin{equation}\label{z-mobius}
|z|^2 = \left| \frac1{z'}+a \right|^2 = a^2 + \frac{1+2a\phi'_1}{\phi'^2_1+\phi'^2_2}.
\end{equation}
The solution $p'(z')$ to the boundary crossing probability is Eq.~\eqref{vr-solution-2d} with $|z|$ written as Eq.~\eqref{z-mobius} in terms of $\phi'_1$ and $\phi'_2$.  This shows that we know the exact solution for any potential $v'(\phi'_1,\phi'_2)$ which is a function only of the combination $(1+\b k \cdot \bs\phi')/|\bs\phi'|^2$, where $\b k$ is any constant vector in the field space.

There is a subtlety in the choice of the $\pm$ signs.  If both of the circles in the $z$-plane have radius smaller than $a$ (like the red and blue circles of figure~\ref{fig:mobius}), then $p'_\pm(z')=p_\pm(z)$.  But if both have radius greater than $a$ (like the magenta and black circles of figure~\ref{fig:mobius}), then  $p'_\pm(z')=p_\mp(z)$, because the inner and outer boundaries are interchanged under the transformation.  Finally, if the smaller circle has radius smaller than $a$ and the larger one has radius greater than $a$ (like the blue and black circles of figure~\ref{fig:mobius}), then this method is not applicable, since the transformation is not holomorphic throughout the annulus $\Omega$.

\subsubsection{Example: Joukowski transformation}

\begin{figure}
\centering
\includegraphics[width=.4\textwidth, trim=4.5cm 0 4.5cm 0, clip, valign=c]{zplane.pdf}
\includegraphics[width=.55\textwidth, valign=c]{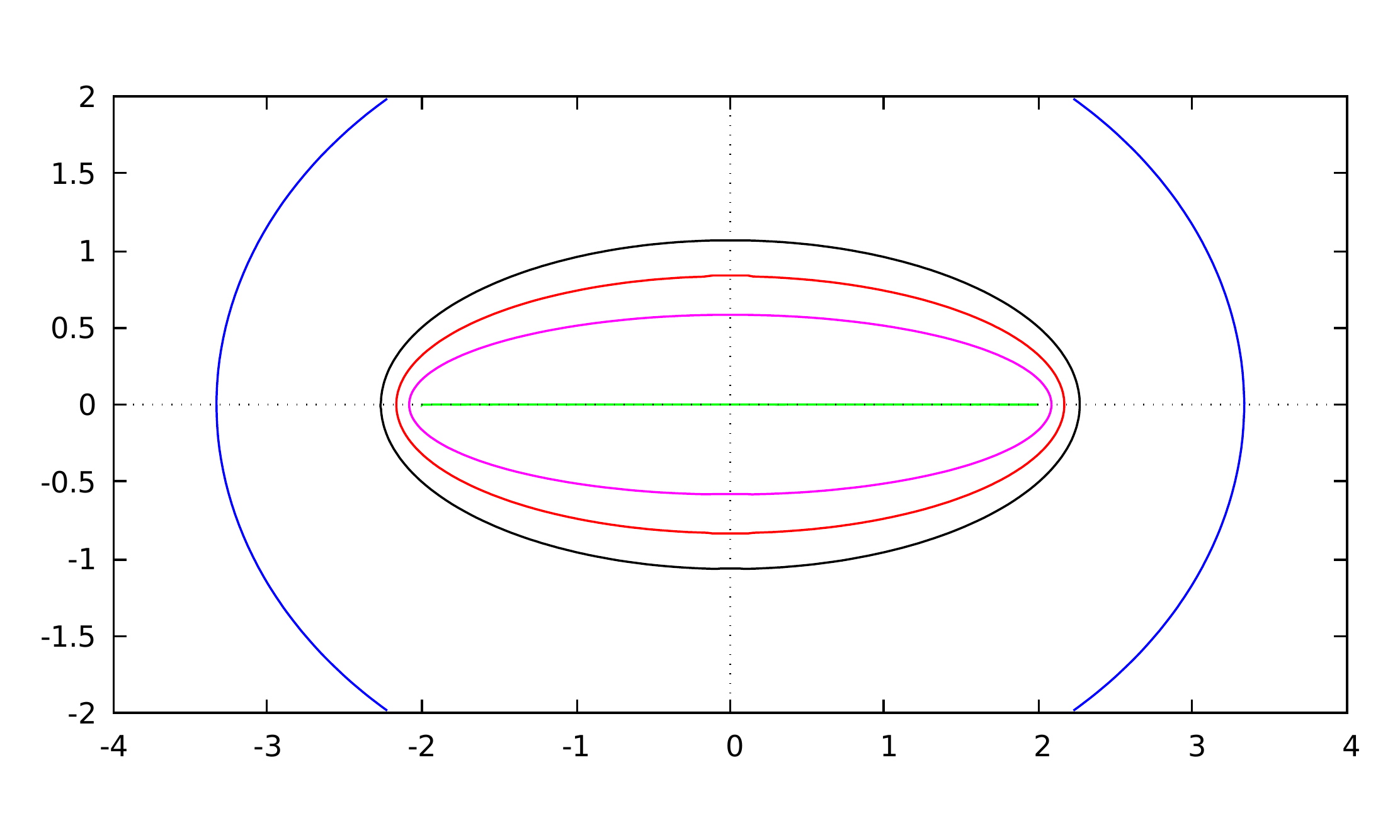}
\caption{The Joukowski transformation $z\mapsto z'=z+1/z$ mapping concentric circles with radii $n/3$ ($n=1,\ldots,5$) in the $z$-plane (left) to their images in the $z'$-plane (right).}
\label{fig:joukowski}
\end{figure}

The Joukowski transformation
\begin{equation}
f(z) = \frac{a+b}{2} \frac{z}{R} + \frac{a-b}{2} \frac{R}{z}
\end{equation}
maps the circle $|z|=R$ to the ellipse with semi-axes $a$ and $b$.  Every other circle $|z|=r>R_c=\sqrt{\frac{|a-b|}{a+b}}R$ is also mapped to an ellipse.  These ellipses fill the entire plane, with the limiting ellipse $|z|=R_c$ mapped to a line segment of length $2\sqrt{|a^2-b^2|}$ (horizontal if $a>b$, and vertical if $a<b$).  The family of circles $|z|=r<R_c$ also gets mapped to ellipses that fill the entire plane, with limiting ellipse $|z|=0$ mapped to infinity.  Thus each ellipse corresponds to two circles with reciprocal radii.  Figure~\ref{fig:joukowski} depicts the transformation for $a=R+1/R$ and $b=R-1/R>0$.

Let us see the effect of the Joukowski transformation on the two-dimensional problem of inflation in the $v(r)$-potential $v(|z|)$.  For $2z'/a = z+1/z$, we can express $|z|$ in terms of the primed fields using
\begin{equation}\label{z-joukowski}
az = z' \pm \sqrt{z'^2-a^2}.
\end{equation}
The solution  $p'(z')$ to the boundary crossing probability is Eq.~\eqref{vr-solution-2d} with $|z|$ written using Eq.~\eqref{z-joukowski} in terms of $\phi'_1$ and $\phi'_2$.  This shows that we know the exact solution for any potential $v'(\phi'_1,\phi'_2)$ which is a function only of the combination $|z' \pm \sqrt{z'^2-a^2}|$, where $a$ is any real number.  In the face of it, this is not a simple class of potentials; it is artificially cooked to map to an exact solution.  But the fact that the contours of constant $v'$ are ellipses makes it a bit more appreciable, i.e., we have managed to solve the problem of boundary crossing probability on any potential whose equipotential contours are of the form of the ellipses described above.

\subsection{Electrostatics Analogy}\label{ssec:electro-analogy}

We saw earlier in subsection~\ref{ssec:d>2} that the master equation~\eqref{master} can be written as
\begin{equation}
\partial_i \left( e^{1/v} \partial_i p \right) = 0.
\end{equation}
This is the statement of vanishing of the divergence $\bs\nabla \cdot \b E$ of the 2-dimensional vector field $\b E = e^{1/v} \bs\nabla p$.  However, this is not enough information to determine $\b E$ (although it is enough to determine $p$ for a given $v$).  So we may demand an extra equation $\bs\nabla \times \b E=0$ and hope that a consistent solution exists.  The pair of constraints $\bs\nabla \cdot \b E = \bs\nabla \times \b E = 0$ is enough (together with appropriate boundary conditions on $\b E$) to determine $\b E$ on $\Omega$.  This reduces the problem of finding the boundary crossing probability to an electrostatics problem in empty space.

It is easy to see that if $f(z) = U(\phi_1,\phi_2) + iV(\phi_1,\phi_2)$ is a holomorphic function on the domain $\Omega$, then the Cauchy-Riemann equations imply that $\b E = (U,-V)$ has vanishing divergence and curl.  In addition, the electrostatic potential $\Phi$ corresponding to the electric field $\b E$ is found to be 
\begin{equation}
\Phi(z) = -\int_{z_0}^z \b E \cdot d\bs\ell = -\operatorname{Re} \int_{z_0}^z f(z')dz',
\end{equation}
where the integration is along any path that connects the reference point $z_0$ to $z$.

The requirement $\bs\nabla \times \b E=0$ implies that $\bs\nabla v \times \bs\nabla p = 0$.  Together with $\b E = -\bs\nabla \Phi$ this means that $\bs\nabla v \parallel \bs\nabla \Phi \parallel \bs\nabla p$.  In other words, the level contours of the three functions $v$, $\Phi$ and $p$ coincide.  Therefore, both $v$ and $p$ can be regarded as functions of $\Phi$ alone.  So we may write $-\bs\nabla \Phi = e^{1/v} \bs\nabla p$ as $-d\Phi = e^{1/v(\Phi)} dp$, which has the solution
\begin{equation}\label{p-vs-Phi}
p = -\int e^{-1/v(\Phi)} d\Phi.
\end{equation}

These results suggest the following strategy to generate/obtain a solution for the master equation.  Given $v$, find an electrostatic potential $\Phi$ (satisfying $\nabla^2\Phi=0$) whose equipotential contours coincide with those of $v$.  Then express $v$ as a function $v(\Phi)$ and plug it into Eq.~\eqref{p-vs-Phi} to obtain the boundary crossing probability $p$.  Since we have a freedom of rescaling on $\Phi$, and since we also have the constant of integration in Eq.~\eqref{p-vs-Phi} at our disposal, we can arrange for $p=0$ and $p=1$ to occur on any of the level contours that we wish.

For a generic $v(\phi)$, an electrostatic potential $\Phi$ with identical equipotential contours may not exist.  If it does, then there must exist a function $\mu$ such that $\bs\nabla \Phi = e^\mu \bs\nabla v$.  Taking the curl, we find that $\bs\nabla \mu$ and $\bs\nabla \Phi$ are parallel.  Taking the divergence and noting that $\Phi$ is harmonic, we find that
\begin{equation}
\bs\nabla \mu = \pm \frac{\nabla^2 v}{\| \bs\nabla v \|^2} \bs\nabla v.
\end{equation}
Since $v$ and $p$, and now $\mu$, are each a function of $\Phi$ alone, the above equation implies that the coefficient
\begin{equation}
g = \frac{\nabla^2 v}{\| \bs\nabla v \|^2}
\end{equation}
must be a function of $v$ alone.  This is precisely the definition of harmonic potentials presented in Ref.~\cite{Assadullahi:2016gkk} (beware that a harmonic potential is not a harmonic function).  Indeed it should have been obvious that this method applies only to harmonic potentials, since they are the ones for which $p$ is constant on level contours of $v$.  Once $g(v)$ is calculated, we can readily find $\mu = \pm \int g(v)dv$ and obtain $\Phi$ from $d\Phi = e^{\mu(v)} dv$.

\subsubsection{Example: $v(r)$-potentials}

As an example, the whole class of $v(r)$-potentials can be derived from the holomorphic function $f=c/z$, which corresponds to $\Phi=-c\log (r/r_0)$, i.e., a point charge located at the origin.  To do so, we need to choose the lower limit of integration for $p_+$ in Eq.~\eqref{p-vs-Phi} to be $r_-$, and choose the constant $c$ to be equal to $1/\int_{r_-}^{r_+} e^{-1/v}\frac{dr}{r}$.  Then we recover Eq.~\eqref{vr-solution-2d}.  This is very interesting and gives us a unifying picture for all $v(r)$-potentials in terms of the electrostatics of a single point charge.

\subsubsection{Example: double-well potential}

\begin{figure}
\centering
\includegraphics[width=.9\textwidth]{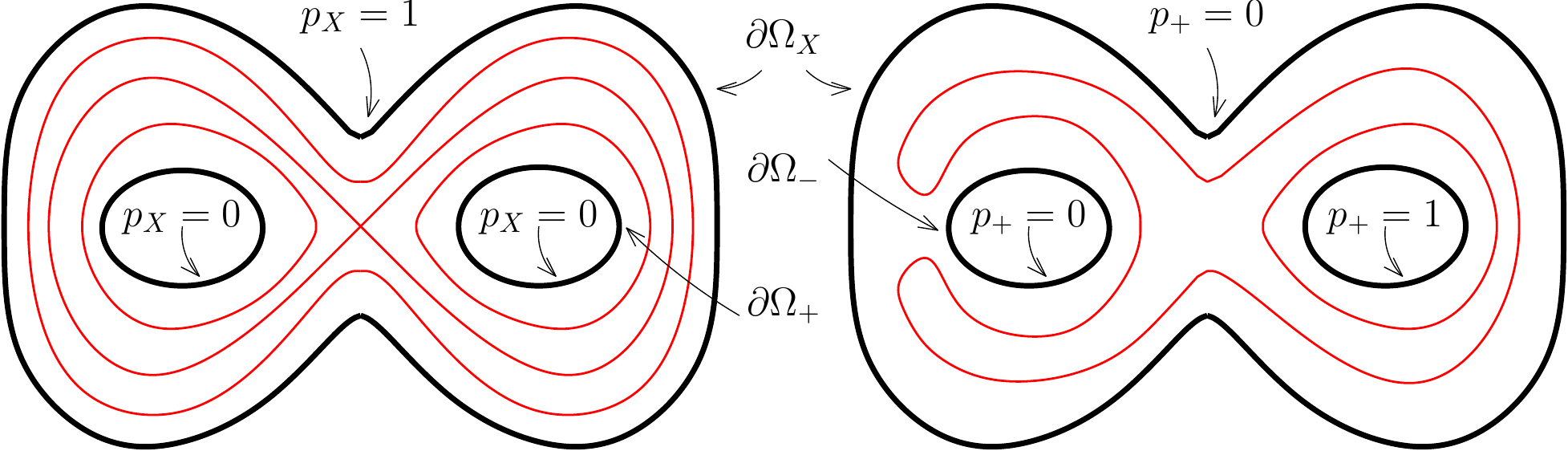}
\caption{Schematic shape of the $p=\text{const}$ contours of the double-well potential, for the boundary crossing probabilities $p_X$ (left) and $p_+$ (right).  In the left panel the equipotential contours $v=\text{const}$ coincide with the $p=\text{const}$ contours.  The equipotential contours of the right panel are the same as the left panel but are not shown.}
\label{fig:double-well}
\end{figure}

As a second example, consider the double-well potential
\begin{equation}
v = v_0 |z-z_+|^2 |z-z_-|^2.
\end{equation}
The equipotential contours of $v$ resemble those of two point charges at $z_\pm$  (see Figure~\ref{fig:double-well}).  This suggests the electrostatic potential
\begin{equation}
\Phi = -q \log |z-z_+| - q \log|z-z_-| + \Phi_0 = -\frac{q}{2} \log \frac{v}{v_0} + \Phi_0.
\end{equation}
The fact that $\Phi$ is a function of $v$ indicates that $v$ is a harmonic potential (equivalently, the function $e^\mu = -q/2v$ exists such that $d\Phi=e^\mu dv$).  We can now use \eqref{p-vs-Phi} to obtain
\begin{equation}
p(v) = \frac{q}{2} \int_{\bar v}^v e^{-1/v'} \frac{dv'}{v'}.
\end{equation}
The two constants $\bar v$ and $q$ must be chosen to satisfy the boundary conditions.

The double-well potential is peculiar in that it requires three boundaries.  Two of the boundaries, which we continue to call $\partial\Omega_\pm$, are the reheating surfaces $v=v_\pm$ surrounding the minima $z_\pm$.  The third boundary, which we call $\partial\Omega_X$ corresponds to a high energy cutoff $v=v_X$ beyond which we do not trust our theory.  The appropriate boundary conditions for $p_\pm$ are just Eq.~\eqref{boundary-conditions} \textit{and} $p_\pm(\partial\Omega_X)=0$.  On the other hand, the boundary conditions for $p_X$ are $p_X(\partial\Omega_\pm)=0$ and $p_X(\partial\Omega_X)=1$.  Unfortunately, we can use this method to compute only $p_X$, but not $p_\pm$.  The reason is that the solution $p(v)$ we find by this method is a function of $v$ alone, but $\partial\Omega_\pm$ both have the same $v$ while $p_+(\partial\Omega_+) = 1 \neq 0 = p_+(\partial\Omega_-)$.  For the same reason, we may calculate $p_X$ only if the potential has the same value on $\partial\Omega_\pm$.

To calculate $p_X$, we can simply choose the constant $\bar v$ to be equal to $v_+=v_-$ (the common value of the potential on $\partial\Omega_\pm$), so that the boundary conditions on $\partial\Omega_\pm$ are satisfied.  The other constant, $q$, is then set to make the normalization right, thus:
\begin{equation}
p_X(v) = \frac{\displaystyle \int_{v_+}^v e^{-1/v'} \frac{dv'}{v'}}{\displaystyle \int_{v_+}^{v_X} e^{-1/v'} \frac{dv'}{v'}},
\end{equation}
where $v_X$ is the value of the potential on $\partial\Omega_X$.  Due to the exponential dependence of the integrand on the potential, we can approximate it as
\begin{equation}
p_X(v) \approx \frac{v}{v_X} \exp \left( \frac{1}{v_X} - \frac1v \right).
\end{equation}
Evidently $p_X$, the probability to escape reheating in the valleys, decays exponentially as the starting point of the inflaton moves away from $\partial\Omega_X$.

\subsubsection{Example: multi-well potential}

\begin{figure}
\centering
\includegraphics[width=.8\textwidth]{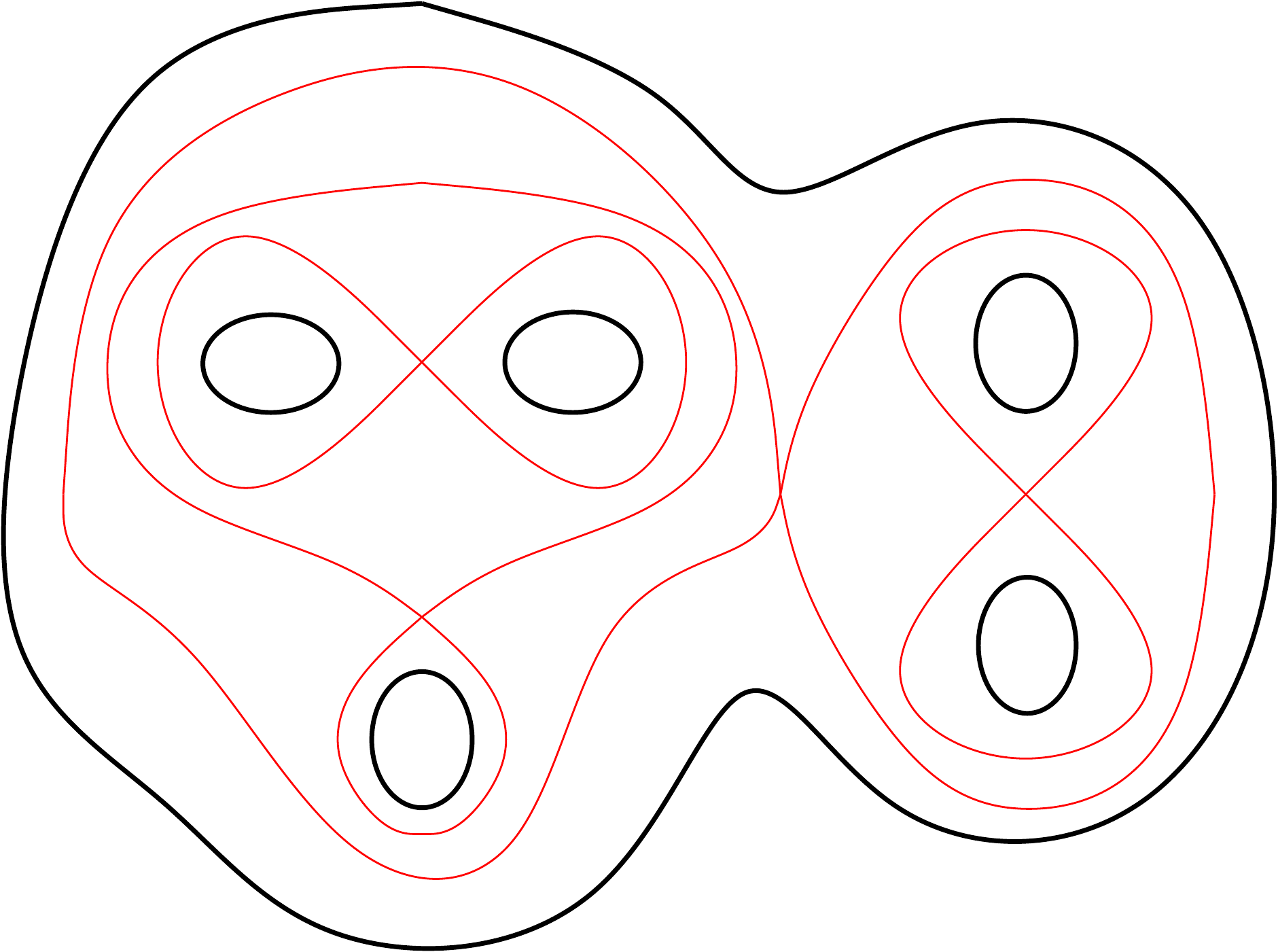}
\caption{The equipotential contours $v=\text{const}$ of a potential with five minima.}
\label{fig:multi-well}
\end{figure}

The preceding result is applicable to a multi-well potential
\begin{equation}\label{multi-well}
v = v_0 \prod_i |z-z_i|^2,
\end{equation}
like Fig.~\ref{fig:multi-well} if, as before, we choose the boundaries $\partial_i\Omega$ around each valley to be at a common value $v_i$ for all $i$.  The electrostatic potential will be
\begin{equation}
\Phi = -q \sum_i \log |z-z_i| + \Phi_0 = -\frac{q}{2} \log \frac{v}{v_0} + \Phi_0,
\end{equation}
so we continue to have
\begin{equation}
p_X(v) = \frac{\displaystyle \int_{v_i}^v e^{-1/v'} \frac{dv'}{v'}}{\displaystyle \int_{v_i}^{v_X} e^{-1/v'} \frac{dv'}{v'}},
\end{equation}
Interestingly, the probability $p_X$ of escaping reheating is independent of the number of the minima in a landscape described by the multi-well potential~\eqref{multi-well}.

\subsection{Other Electromagnetic Analogies}\label{ssec:other-analogies}

In subsection~\ref{ssec:electro-analogy} we identified $e^{1/v} \bs\nabla p$ as the electric field $\b E$ and demanded a supplementary equation $\bs\nabla\times\b E=0$ in addition to the master equation $\bs\nabla\cdot\b E=0$.  Although this proved fruitful in obtaining new solution, we saw that the supplementary constraint restricts the solvable problems to harmonic potentials.  We can make other less restrictive analogies that work in more general situations, although they are not as fruitful as the previous one.

\subsubsection{Magnetostatic Analogy}

Since the master equation~\eqref{master} asserts that the divergence of $e^{1/v} \bs\nabla p$ is zero, we can identify it as the magnetic field
\begin{equation}
\b B = e^{1/v} \bs\nabla p.
\end{equation}
Its curl may then be specified by a current density $\b J$,
\begin{equation}
\bs\nabla \times \b B = \b J.
\end{equation}
All of the previous examples in subsection~\ref{ssec:electro-analogy} can be translated by replacing electric point charges with magnetic point charges and setting $\b J=0$.  More general situations correspond to nonzero electric current $\b J\neq 0$.  Nonetheless, this is not a very helpful analogy, since the magnetic field lines of a current are closed loops.  The contours of constant $p$, which are perpendicular to $\bs\nabla p \parallel \b B$, therefore resemble rays emanating from a point.  However, as we saw in the physically interesting cases, we have contours of $p=\text{const}$ that are closed loops.

\subsubsection{Dielectric Medium Analogy}

Let us regard $\Phi=-p$ as the electrostatic potential, so that $\b E = \bs\nabla p$, and identify
\begin{equation}
\b D = e^{1/v} \bs\nabla p
\end{equation}
as the displacement current.  Comparing with $\b D=\epsilon \b E$, it is clear that in this analogy $e^{1/v}>1$ plays the role of the permittivity $\epsilon$ and is variable across the dielectric medium (note that $\epsilon_0=1$ in our convention).  According to the master equation~\eqref{master}, $\bs\nabla\cdot\b D=0$, which means that there is no free charge density, $\rho_f=0$.  Of course, the total charge density is nonzero and is given by $\rho = \bs\nabla\cdot\b E = -\bs\nabla \frac1v \cdot \bs\nabla p$, but it is comprised entirely of the bound charges induced in the dielectric medium. 

With this analogy, every boundary crossing problem for an arbitrary potential $v$ on an arbitrary boundary $\partial\Omega$ can be cast into a well-defined electrostatic problem: The electrostatic potential $\Phi$ is specified on the conducting surfaces $\partial\Omega$ in a medium with inhomogeneous permittivity $\epsilon=e^{1/v}$; find $\Phi$ everywhere in the medium.  In principle, the problem of finding $p_\pm$ in the double-well potential which was impossible to solve with the method of subsection~\ref{ssec:electro-analogy} can be expressed as a problem in such a dielectric medium.  Of course, in general such problems are hard to solve analytically, as we don't know the total charge distribution in advance.  But all numerical methods of classical electrostatics may be applied to our problems.

\section{Summary and Discussion}\label{sec:summary}

In the first part of this paper, we generalized the proof of the criticality of $D=2$ to arbitrary shapes of potentials, a result we dubbed ``$v(\phi)$-criticality'' to distinguish from the previously known result~\cite{Assadullahi:2016gkk}, ``$v(r)$-criticality'', which was proved only for $v(r)$-potentials.  The new statement is that the probability for the inflaton to escape to infinitely far regions in the field space is identically zero in $D\leq2$ dimensions.  This is in line with P\'olya's famous recurrence theorem~\cite{Polya}, which states that random walks are recurrent on $D=1,2$-dimensional lattices (i.e., the walker will eventually return to its starting point with probability 1).\footnote{Or as the saying goes: ``A drunk man will find his way home, but a drunk bird may get lost forever.''}  Of course, there are distinctions:  Our setup is on a continuous space while P\'olya's theorem is about discrete lattices; and we talk about the probability of avoiding escape to infinity, rather than returning to the exact initial position.  Nevertheless, there is considerable similarity between the two results.

In the second part, we developed several methods for calculating the boundary crossing probability in two dimensions.  We used conformal transformations to generate problems for which we can find an analytic solution, e.g., problems with elliptical equipotential contours.  We also drew an analogy with electrostatics problems and, as a result, we were able to obtain the probability of escaping reheating for a double-well inflaton potential in $D=2$ dimensions.  We observed the exponential decay characteristic of such situations.  Most notably, we did the same for a multi-well potential and we showed that the probability of escaping reheating is independent of the number of wells.  This is relevant for considering inflation on a landscape and studying when and with what probability inflation will end.  The remarkable fact that this probability is independent of the number of minima may be a peculiar feature of the multi-well potential we used, but it interesting on its own and deserves further consideration.

Finally, note that, except for the conformal transformation method, the other methods of section~\ref{sec:2d} are actually applicable to higher dimensions as well.  Nevertheless, we concentrated only on examples in $D=2$.

\section{Acknowledgements}

We would like to thank Hooshyar Assadullahi, Vincent Vennin and David Wands for discussions in earlier collaborations.   M.N.\ is thankful to Mohammad Safdari and Abbas~Ali Saberi for helpful discussions and explanations.  M.N.\ also acknowledges financial support from the research council of University of Tehran.

\appendix

\section{Harmonic Functions on Annular Regions}\label{app:harmonic}

In this appendix we solve Laplace's equation on an annular region $\Omega = \{ \phi | r_- \leq r \leq r_+ \}$ of the flat space in the limit that $r_+$ goes to infinity.  The point is to illustrate the different behavior of $D>2$ and $D\leq2$; specifically, that the latter does not admit a proper solution when boundary conditions are specified on both $\partial\Omega_-$ and $\partial\Omega_+$.

Let us consider $D>2$ first.  In $\mathbb{R}^D$ and in polar coordinates, the metric reads $ds^2 = dr^2 + r^2 d\Omega^2$, where $d\Omega^2$ is the metric on the unit sphere $S^{D-1}$.  Let us denote the components of the metric by $g_{rr}=1$ and $g_{\theta_i\theta_i} = r^2 \gamma_{ii}$, where $\gamma_{ii}$ depends only on the angular coordinates $\theta_i$ of $S^{D-1}$ but not on $r$.  The metric determinant is $g = r^{2(D-1)} \gamma$.  Therefore,
\begin{equation}
\begin{aligned}
\nabla^2 p &= \frac{1}{\sqrt g} \partial_a \left( \sqrt g g^{ab} \partial_b p \right) \\
&= \partial^2_r p + \frac{D-1}{r} \partial_r p + \sum_{i=1}^{D-1} \frac{1}{r^2 \sqrt \gamma} \partial_i \left( \sqrt \gamma \gamma^{ii} \partial_i p \right).
\end{aligned}
\end{equation}
We can now use separation of variables and write $p(r,\theta) = R(r) \Theta(\theta)$.  Then
\begin{equation}
\frac{\nabla^2 p}{p} = \frac{R''}{R} + \frac{D-1}{r} \frac{R'}{R} + \frac{1}{r^2} \sum_{i=1}^{D-1} \frac{1}{\sqrt \gamma \Theta} \partial_i \left( \sqrt \gamma \gamma^{ii} \partial_i \Theta \right).
\end{equation}
It follows immediately that $\Theta$ must be an eigenfunction of the Laplacian on $S^{D-1}$,
\begin{equation}
\sum_{i=1}^{D-1} \frac{1}{\sqrt \gamma} \partial_i \left( \sqrt \gamma \gamma^{ii} \partial_i \Theta \right) = k \Theta,
\end{equation}
and $R=r^s$, where $s(s+D-2)=-k$.  It is known that the eigenvalues $k$ of the Laplacian are of the form $k=-l_1(l_1+D-2)$ where $l_1\geq0$ is an integer, in order for the corresponding eigenfunctions $\Theta = Y_{l_1,\ldots,l_{D-1}}(\theta)$ to be regular (these are a generalization of the classical spherical harmonics; the $D-2$ integer indices are akin to the $m$ in $Y_{lm}$ and satisfy $l_1 \geq l_2 \geq \ldots \geq l_{D-2} \geq |l_{D-1}|$).  The $Y_{l_1,\ldots,l_{D-1}}$s form a complete basis for the functions on $S^{D-1}$ and the final answer is
\begin{equation}\label{Y-series}
p(r,\theta) = \sum_{l_1,\ldots,l_{D-1}} \left( a_{l_1,\ldots,l_{D-1}} r^{l_1} + b_{l_1,\ldots,l_{D-1}} r^{-l_1-D+2} \right) Y_{l_1,\ldots,l_{D-1}}(\theta),
\end{equation}
where $a$ and $b$ are constants to be determined by boundary conditions.

A similar approach works for $D=2$, except that when $l_1=0$, the second solution is not $r^{-l_1}$, rather $\log r$.  Then we have, instead of Eq.~\eqref{Y-series}:
\begin{equation}\label{Y-series=D=2}
p(r,\theta) = a_0 + b_0 \log r + \sum_{l=1}^\infty \left( a_{l} r^{l} + b_{l} r^{-l} \right) \cos(l\theta).
\end{equation}
Finally, the case of $D=1$ is trivial, giving
\begin{equation}\label{Y-series-D=1}
p(r) = a_0 + b_0 r.
\end{equation}

The purpose of deriving Eqs.~\eqref{Y-series}, \eqref{Y-series-D=1} and \eqref{Y-series=D=2} is to illustrate the role of $D$ in the existence of solutions on unbounded domain (domains that extend to $r=\infty$).  For concreteness, we consider the domain $r_-\leq r\leq r_+$ in the limit $r_+\to\infty$.  Let us first consider $D>2$.  As $r\to\infty$, the second term in~\eqref{Y-series} vanishes no matter what $l_1$ or $b$'s are.  In order to have a finite value for $p(r_+)$, the first term must vanish for all $l_1>0$, i.e., $a_{l_1,\ldots,l_{D-1}} = 0$ unless $l_1=0$.  Noting that $Y_{l_1,\ldots,l_{D-1}}$ is a constant for $l_1=0$, we conclude that the only possible bounded solutions on an unbounded domain are those that approach a constant at $r=\infty$.  On the other hand,
\begin{equation}\label{p-at-R-}
p(r_-) = \left( a_{0,\ldots} + b_{0,\ldots} \right) Y_{0,\ldots} + \sum_{l_1>0} b_{l_1,\ldots,l_{D-1}} r_-^{-l_1-D+2} Y_{l_1,\ldots,l_{D-1}},
\end{equation}
which can have arbitrary $\theta$-dependence.  So for example, in $D=3$, there is no solution of Laplace's equation that approaches the non-uniform profile $Y_{11}$ as $r\to\infty$, whereas it is perfectly possible for $p$ to approach $Y_{11}$ as $r\to r_-$.  It is important to notice that $a_{0,\ldots}$ is not determined from $p(r_-)$ through Eq.~\eqref{p-at-R-}, so the value of $p(r_+=\infty)$ can be specified as an arbitrary constant (which is 0 or 1 for $p_-$ and $p_+$, respectively), independent of $p(r_-)$.

Now consider $D=2$.  The corresponding problem on the domain $r_-\leq r\leq r_+$ with bounded solution must have $a_l=0$ for all $l>0$.  But since $\log r$ is unbounded, $b_0$ must vanish too.  Therefore,
\begin{equation}
p(r_-) = a_0 + \sum_{l=1}^\infty \left( a_{l} r^{l} + b_{l} r^{-l} \right) \cos(l\theta).
\end{equation}
We see that again $p$ can have an arbitrary $\theta$-dependent profile on $\partial\Omega_-$, but it must approach a uniform $\theta$-independent value $a_0$ as $r\to\infty$.  The crucial difference with $D>2$ is that $b_0$ is absent.  Thus, the value of $a_0$ (the boundary condition at infinity) is fixed by the boundary condition $p(r_-)$ at $\partial\Omega_-$.  So once $p(r_-)$ is specified, no further boundary condition can be specified for the $r=\infty$ boundary.  The same thing happens for $D=1$, precisely because $r$ (just like $\log r$) is an unbounded function.

\section{Image Charges in $D=2$ and $D=3$}\label{app:image}

In this appendix, we demonstrate the distinction between the $D=2$ and $D=3$ cases on unbounded domains using the technique of image charges.

Let us begin by $D=3$.  For the sake of simplicity, we choose a constant potential $v$ over the domain $\Omega = \{ \phi | r>r_- \}$.  Since this is a $v(r)$-potential, we can use Eq.~\eqref{vr-solution} with $r_+=\infty$:
\begin{equation}
p_-(\phi) = \displaystyle \dfrac{ \displaystyle \int_{r}^{\infty} r'^{-2} e^{-\frac{1}{v}} d r'}{\displaystyle\int_{r_-}^{\infty} r'^{-2} e^{-\frac{1}{v}} dr'} = \frac{1/r}{1/r_-} = \frac{r_-}{r}.
\end{equation}
Let us see how this comes about using image charges.  According to subsection~\ref{ssec:d>2}, $p_-(\phi)$ is the image charge inside the sphere, due to a negative unit point charge at $\phi$.  We can thus apply the well-known result~\cite{Jackson} that the image charge, due to a point charge $q$ at a distance $r$ away from the center of a grounded sphere of radius $r_-$, is given by $q'=-qr_-/r$.  Setting $q=-1$, we readily recover the result above.

Now consider $D=2$, again with constant $v$ and on the domain $\Omega = \{ \phi | r>r_- \}$.  Once again, we can use Eq.~\eqref{vr-solution} with $r_+=\infty$:
\begin{equation}
p_-(\phi) = \lim_{r_+\to\infty} \displaystyle \dfrac{ \displaystyle \int_{r}^{r_+} r'^{-1} e^{-\frac{1}{v}} d r'}{\displaystyle\int_{r_-}^{r_+} r'^{-1} e^{-\frac{1}{v}} dr'} = \lim_{r_+\to\infty} \frac{\log(r_+/r)}{\log(r_+/r_-)} = 1.
\end{equation}
This time the problem corresponds to a line of charge (along the $z$-axis in 3 dimensions) lying outside a grounded cylinder with parallel axis.  The image charge will be another line of charge inside the cylinder, with an opposite but \textit{equal} linear charge density~\cite{Jackson}. 
Thus $p_-(\phi)$, being the ratio of the image charge to the original charge, is equal to one, regardless of the distance of the original charge from the cylinder.

It is important to contrast the two cases.  In $D=3$, $p_-$ decays like $1/r$, while in $D=2$, $p_-$ is constant.  The crucial difference is that, as the original charge moves away from the grounded surface, its potential on the surface decreases for $D=3$, but not for $D=2$.  Therefore, the interior image charge $p_-$ required to compensate for this potential decreases in $D=3$, but remains constant in $D=2$.

\end{document}